\documentclass[onecolumn]{revtex4-2}
\usepackage{graphicx} 
\usepackage{amssymb}
\usepackage{amsmath}
\usepackage[colorlinks=true]{hyperref}
\usepackage{ulem}
\usepackage{float}
\usepackage{color}
\usepackage{xcolor}

\begin{document}

\title{On Avoidance of PBH overproduction in inflation model with modified dispersion relation II: the Peak Theory Approach}

\author{Chengrui Yang}
\email{ycrd@hust.edu.cn}
\affiliation{School of Physics, Huazhong University of Science and Technology, Wuhan, 430074, China}

\author{Weixin Cai}
\email{weixincai@hust.edu.cn}
\affiliation{School of Physics, Huazhong University of Science and Technology, Wuhan, 430074, China}

\author{Taotao Qiu}
\email{qiutt@hust.edu.cn}
\thanks{Corresponding author.}
\affiliation{School of Physics, Huazhong University of Science and Technology, Wuhan, 430074, China}

\begin{abstract}
The inflation model, where the dispersion relation of the inflaton field gets modified, can generate primordial black holes as well as scalar-induced gravitational waves at small scales. However, the literature shows that when taking the recent observational data from pulsar timing arrays into account, the ``overproduction problem" for primordial black holes will appear for some inflation models. In our previous paper, we calculated the abundance of primordial black holes in our model using the compaction-function approach, and found that the overproduction issue can get alleviated with and without non-Gaussianity effects. In this paper, we revisit this problem using another approach, i.e., the peak theory. We find that the peak theory approach makes the situation worse, thus we cannot avoid overproduction without the help of non-Gaussianity. However, with a small negative non-Gaussianity of $f_{\mathrm{nl}}\leq -1$, the problem can be fixed. Moreover, since the primordial black holes generated generally have nearly solar masses, we also consider the constraints from microlensing observations, and see that our results fit the data very well.
\end{abstract}

\maketitle

\section{Introduction}
The Primordial Black Hole (PBH) \cite{Zeldovich:1967lct, Hawking:1971ei, Carr:1974nx} is a kind of interesting object in the fields of gravity, cosmology and astrophysics. It can not only serve as one of the dark matter candidates, but also generate gravitational waves and/or Hawking radiation which can be observed via various approaches, therefore serving as a useful probe of the early universe. There have been wide investigations of PBHs from various aspects with many interesting achievements, see \cite{Khlopov:2008qy, Yuan:2021qgz, Villanueva-Domingo:2021spv, Oncins:2022ydg, Escriva:2022duf, Carr:2025kdk, Shankaranarayanan:2026hnn} for comprehensive reviews.

It is fundamental to ask how the PBHs can be generated. As is well known, the main pathway of generating PBHs is the enhancement of primordial curvature perturbations at the end of the early time inflation, and one of the possibilities to do so is to suppress the sound speed $c_s$ of the inflaton field. However, due to the potential strong coupling problem \cite{Cheung:2007st, Ballesteros:2018wlw}, one has to introduce some correction terms such as $k^4$ term to the dispersion relation of the curvature perturbation during inflation, make it modified as $\omega^2=c_s^2k^2+\alpha_M^2k^4$. Such kind of model can be dubbed as ``modified dispersion relation (MDR) model" \cite{Arkani-Hamed:2003juy,Arkani-Hamed:2003pdi,Ballesteros:2021fsp}. The generation of PBHs within such a MDR model has been well investigated, see \cite{Ballesteros:2018wlw, Ballesteros:2021fsp, Gorji:2021isn, Qiu:2022klm, Yang:2025vsq}.

Recently, several groups of pulsar timing array (PTA) observations have released their latest data, which contain signals of gravitational waves in the nanohertz frequency band \cite{NANOGrav:2023gor, NANOGrav:2023hvm, Reardon:2023gzh, Zic:2023gta, EPTA:2023fyk, EPTA:2023gyr, Xu:2023wog}. These signals present the Hellings--Downs correlation pattern, which supports a stochastic gravitational wave background (SGWB) at the $2$--$4\sigma$ level. There are various possible candidates for such kind of gravitational waves, see \cite{Ellis:2023dgf, Grishchuk:1974ny, Starobinsky:1979ty, Tomita:1967wkp, Matarrese:1993zf, Matarrese:1997ay, Acquaviva:2002ud, Ananda:2006af, Domenech:2021ztg, Witten:1984rs, Hogan:1986dsh, Vilenkin:1984ib, Burden:1985md, Hindmarsh:1994re, Ellis:2023oxs}. Among these candidates, the gravitational wave induced by the scalar perturbation during the inflation period is of particular interest, since it not only has better fit to the data with model comparison Bayesian analysis \cite{Figueroa:2023zhu}, but can also be connected with one of the most promising dark matter candidates: the primordial black holes. However, later investigations have found that, in order for the scalar-induced gravitational waves (SIGW) to be aligned with the PTA data, the abundance of accompanying PBHs may exceed $100\%$, which is theoretically not possible \cite{Franciolini:2023pbf, Dandoy:2023jot}. Such a contradiction is called the overproduction problem.

Several approaches have been proposed. For example, a negative non-Gaussianity can suppress the density fluctuations that collapse into PBHs, thereby reducing their abundance \cite{Franciolini:2023pbf, Wang:2023ost, Liu:2023ymk, DeLuca:2023tun, Firouzjahi:2023xke, Chang:2023aba, Pi:2024lsu, Inui:2024fgk}. Other ways such as modification of the window function/threshold are also useful \cite{Inomata:2023zup}. It is also possible that the PBHs were generated not in the traditional radiation-dominated era, but an era with a different equation of state of the universe \cite{Domenech:2019quo, Liu:2023pau, Liu:2023hpw, Choudhury:2023fjs}. Actually, since it is well-known that PBHs are generated from primordial fluctuations on small scales where the scalar power spectrum gets enlarged, modifications on small scales may affect the PBH abundance. In our previous paper \cite{Yang:2025vsq}, we investigated whether the MDR inflation models could play a role in reducing the PBH overproduction. Under such corrections, the scalar power spectrum will be enhanced to form a broken-power-law shape. Using the compaction-function approach \cite{Shibata:1999zs}, we calculated the abundance of PBHs formed in this case and found that the overproduction problem can be alleviated. Moreover, with negative non-Gaussianity estimator $f_{\mathrm{nl}}\simeq -1$, the abundance of PBHs is well consistent ($<2\sigma$) with the PTA constraint on SIGW.

However, there are actually various approaches to calculating the PBH abundance in the market, and which approach is more precise is still being debated. Besides the compaction function one, another famous approach is called peak theory \cite{Bardeen:1985tr}. Instead of the traditional Press--Schechter formalism which considers the density contrast inside the horizon exceeding the threshold value as the criterion of PBH formation, the peak theory assumes that PBHs are only generated where the curvature perturbation reaches the top of a peak and counts the peaks above the threshold inside the horizon \cite{Bardeen:1985tr, Yoo:2022mzl, Young:2024jsu}.
Peak theory has been applied in many cases of PBH abundance calculations. See \cite{Yoo:2019pma, Riccardi:2021rlf, Kitajima:2021fpq, Abe:2022xur, Inui:2024fgk} for the effect of local-type non-Gaussianities (especially of the exponential-tail distribution) on the PBH abundances. See \cite{Yoo:2020dkz, Pi:2024ert} for consideration of broad curvature spectrum. See \cite{Mahbub:2021qeo, Wang:2021kbh, Yi:2022anu} for PBH formation in scalar-driven inflation models. See \cite{Banerjee:2022xft, Papanikolaou:2024fzf} for bounce scenarios. See \cite{Saito:2023fpt, Saito:2024hlj, Ye:2025wif, Mohammadi:2025avz} for consideration of PBH spins. See \cite{Escriva:2022duf} for a comprehensive review.

The difference in perspective between the Press--Schechter formalism and the peak theory may result in different results in PBH formation in one model. In order to make a complete exploration on the PBH formation process in various formalisms and compare their differences, in this paper, we will reinvestigate the PBH abundance and the overproduction avoidance of the inflation model with modified dispersion relation, using the peak theory approach. The rest of the paper is arranged as follows: in Sec.~\ref{model} we briefly review the model, with scalar power spectrum, SIGW and PBH abundance calculated in Ref.~\cite{Yang:2025vsq}; in Sec.~\ref{peak} we reanalyze the PBH abundance in peak theory, and compare the results with those from the compaction-function approach. Sec.~\ref{conclusion} is the conclusive remarks.

\section{The Model}
\label{model}

\subsection{Curvature Perturbation and Primordial Power Spectrum}


Same as in Ref.~\cite{Yang:2025vsq}, we consider the inflation model where the dispersion relation of the inflaton perturbation is modified with a $k^4$ correction. The Mukhanov--Sasaki equation of the curvature perturbation can be generally written as:
\begin{equation}
\label{zetabar}
{u_k''} {\mathrm{ + }}\left( c_s^2{k^2} + {{\alpha}_M^2}{k^4}{\tau ^2} - \frac{z^{\prime\prime}}{z} \right){u _k} = 0~.
\end{equation}
Equation \eqref{zetabar} describes the evolution of the curvature perturbation mode $u_k\equiv z\zeta$, where $\zeta$ is the curvature perturbation. In this equation, we denote $c_s$ as the sound speed, $\alpha_M$ as the coefficient of the $k^4$ correction term, $\tau$ as the conformal time, and $z$ as a variable related to the background evolution with $z\propto a(\tau)$. Note that although we are discussing a general equation of motion instead of specific models, Eq.~\eqref{zetabar} can be applied in various models. For example, a ghost-condensate model containing a higher-derivative term has been studied in Ref.~\cite{Gorji:2021isn}, whose equation of motion is the same as \eqref{zetabar} \cite{Yang:2025vsq}. Moreover, this equation can also be embedded in the effective field theory (EFT) approach. It has been demonstrated in Ref.~\cite{Cai:2016thi, Cai:2017tku} (see also \cite{Ballesteros:2021fsp}) that for the EFT action:
\begin{eqnarray}
    S&=&\int \mathrm{d}^4x\sqrt{-g}\Bigg[\frac{M_p^2}{2}f(t)R-\Lambda(t)-c(t)g^{00}\nonumber\\
   && +\frac{M_2^4(t)}{2}(\delta g^{00})^2-\frac{m_3^3(t)}{2}\delta K\delta g^{00}-m_4^2(t)(\delta K^2-\delta K_{\mu\nu}\delta K^{\mu\nu})+\frac{\tilde{m}_4^2}{2}R^{(3)}\delta g^{00}\nonumber\\
  &&  -\bar{m}_4^2(t)\delta K^2+\frac{\bar{m}_5(t)}{2}R^{(3)}\delta K+\cdots \Bigg]~,
\end{eqnarray}
as long as the parameters $\bar{m}_4(t)$ or $\bar{m}_5(t)$ do not vanish, a $k^4$ higher-order correction term naturally comes out in the dispersion relation.

Moreover, we consider the sound speed $c_s$ as:
\begin{equation}%
\label{soundspeed}
c_s= \left\{ {\begin{array}{*{20}{r}}
c_{s1} \simeq 1~,&\tau<\tau_\ast\\
c_{s2} \ll 1~,&\tau>\tau_\ast
\end{array}} \right.
\end{equation}
which separates the whole inflationary process into two regions as ``slow-roll I phase" and ``slow-roll II phase" respectively. While in the earlier ``slow-roll I phase" where large-scale perturbations exit the horizon, the inflation occurs as a normal slow-rolling one, in the later ``slow-roll II phase" where small-scale perturbations exit the horizon, the $k^4$ term becomes significant and dominates Eq.~\eqref{zetabar}, which will alter the dispersion relation and is expected to have some effect on PBH generation.

The solution of Eq.~\eqref{zetabar} is performed in \cite{Yang:2025vsq}, and here we show the results:
\begin{equation}
\label{curvature}
     u^{\mathrm{srII}}_{k} = \left\{
     \begin{array}{cc}
          - 2{\mathrm{i}}{{\mathrm{\pi }}^{ - 2}}{\mathrm{\Gamma }}\left( {3/4} \right){\mathrm{\Gamma }}\left( {{\mathrm{1}}/{\mathrm{4}}} \right){\mathrm{\Gamma }}\left( {{\mathrm{3}}/{\mathrm{2}}} \right)c_{s1}^{-3/2}{C^{\mathrm{srI}}_1}{\left( { - \tau _\ast} \right)^{ - 1}}{k^{ - 3/2}},&  k<\frac{1}{-c_{s1}{\tau_\ast}}\\
       \frac{4}{3}{2^{ - 3/2}}{{\mathrm{\pi }}^{ - 1/2}} {\mathrm{i}}{{\mathrm{e}}^{ -{{\mathrm{i}} c_{s1}k{\tau _\ast}}}}c_{s1}^{1/2}\tau _\ast^2{C^{\mathrm{srI}}_1}{\left( { - \tau_\ast } \right)^{ - 1}}{k^{1/2}},&  \frac{1}{-c_{s1}{\tau_\ast}}<k<\sqrt{\frac{2}{\alpha_M\tau_\ast^2}} \\
          \sqrt 2 {{\mathrm{\pi }}^{ - 1}}{\mathrm{i}}{{\mathrm{e}}^{-{\mathrm{i}} c_{s1}k{\tau _\ast}}}{\mathrm{\Gamma }}\left( {3/4} \right){\alpha_M ^{ - 5/4}}c_{s1}^{1/2}{\left( { - {\tau _\ast}} \right)^{ - 1/2}}{C^{\mathrm{srI}}_1}{\left( { - \tau_\ast } \right)^{ - 1}}{k^{ - 2}},&k>\sqrt{\frac{2}{\alpha_M\tau_\ast^2}}
     \end{array}\right.
\end{equation}

From the above one can see that the power spectrum $P_\zeta\equiv\frac{k^3}{2\pi^2}\left|\frac{u}{z}\right|^2$ can be well-parametrized in the broken power-law (BPL) form as:
\begin{equation}
\label{BPL}
    {P_{\zeta}}\left( k \right) = \left\{ {\begin{array}{*{20}{c}}
{A{{\left( {\frac{k}{{{k_ * }}}} \right)}^4}}&{ k < {k_ * }}~,\\
{A{{\left( {\frac{k}{{{k_ * }}}} \right)}^{ - 1}}}&{k > {k_ * }}~,
\end{array}} \right.
\end{equation}
This BPL form indicates that for $k < k_\ast$, the spectrum scales as $k^4$, while for $k > k_\ast$, it scales as $k^{-1}$. This behavior is crucial for generating scalar-induced gravitational waves and primordial black holes.

\begin{figure}
    \centering
    \includegraphics[width=0.75\linewidth]{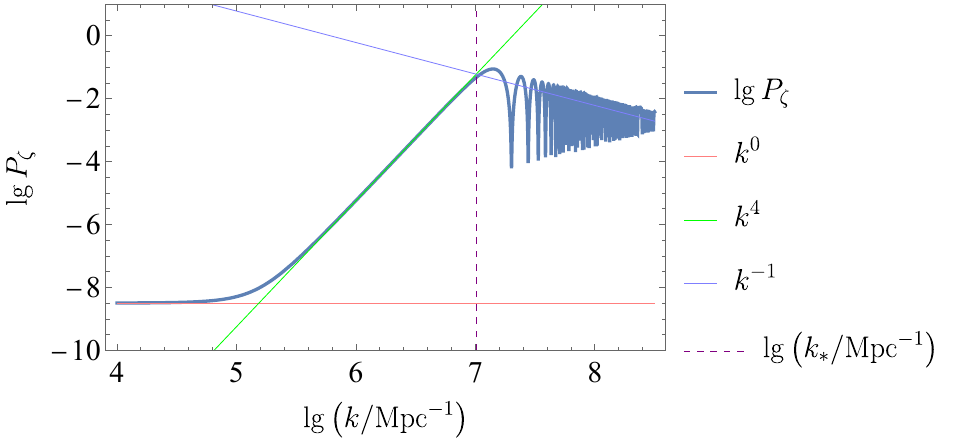}
    \caption{The power spectrum of curvature perturbation obtained from Eq.~\eqref{zetabar} with respect to $k$ (in logarithm). Blue curve represents the accurate result of Eq.~\eqref{curvature}. Three straight lines mimicking the $P_\zeta-k$ relations are also shown for comparison: red line is for $P_\zeta\sim k^0$, green is for $P_\zeta\sim k^4$, while blue is for $P_\zeta\sim k^{-1}$. The dashed line denotes the pivot scale $k_\ast$ corresponding to the peak of the power spectrum.}
    \label{fig:spectrum}
\end{figure}

In order to validate such a parametrization, we numerically calculate the power spectrum and plot it in Fig.~\ref{fig:spectrum}. In the numerical calculation, we assume $c_s=1$ for slow-roll I phase and $c_s=0$ for slow-roll II phase for simplicity. One can see that the parametrization fits well with the numerical results. Moreover, the parameters for amplitude $A$ and for pivot scale $k_\ast$ in Eq.~\eqref{BPL} are related to the model parameters $\alpha_M$ and $\tau_\ast$:
\begin{align}
\label{eq:MDR-BPL-log}
\lg A =\lg\mathcal{P}_\zeta^{\mathrm{CMB}}+\lg\left(c_0^4/9\right)-2\lg\alpha_M\ ,~~~
\lg\left(\frac{k_\ast}{\mathrm{Mpc}^{-1}}\right) =\lg c_0-\tfrac12\lg\alpha_M-\lg\left(\frac{|\tau_\ast|}{\mathrm{Mpc}}\right)\ .
\end{align}
with
\begin{equation}
    c_0\;\equiv\;\left[\frac{9\,\Gamma^2(3/4)}{\pi}\right]^{1/5}\simeq1.34.
\end{equation}

The inverse relationship calls:
\begin{equation}
\label{eq:MDR-BPL-inv}
\lg\alpha_M=-\tfrac12\left[\lg A-\lg\mathcal{P}_\zeta^{\mathrm{CMB}}-\lg\left(c_0^4/9\right)\right]\ ,\qquad
\lg\frac{|\tau_\ast|}{\mathrm{Mpc}}=\lg c_0-\tfrac12\lg\alpha_M-\lg\frac{k_\ast}{\mathrm{Mpc}^{-1}}\ .
\end{equation}
So once $A$ and $k_\ast$ are constrained using the parametrization form \eqref{BPL}, one can apply the relations \eqref{eq:MDR-BPL-log} and \eqref{eq:MDR-BPL-inv} to impose constraints on the model parameters $\alpha_M$ and $\tau_\ast$.

\subsection{Scalar-Induced Gravitational Waves}

The scalar-induced gravitational waves (SIGW) are produced by second-order interactions of curvature perturbations. The tensor power spectrum at horizon reentry is given by \cite{Ananda:2006af, Kohri:2018awv, Domenech:2021ztg}:
\begin{equation}
    P^{\langle k\rangle}_h(k) = 4\int_1^\infty  \mathrm{d}t\int_0^1 \mathrm{d}s{{\left[ {\frac{{\left( {{t^2} - 1} \right)\left( {1 - {s^2}} \right)}}{{{t^2} - {s^2}}}} \right]}^2}{I^2}\left( {t,s} \right){P_\zeta }\left( {k\frac{{t - s}}{2}} \right){P_\zeta }\left( {k\frac{{t + s}}{2}} \right)~,
\end{equation}
where $t\equiv (|{\bf k}-{\bf q}|+q)/k$ and $s\equiv(|{\bf k}-{\bf q}|-q)/k$ are normalized wavenumbers, while $I(t,s)$ is the transfer function that encodes the evolution of perturbations. For the gravitational waves generated in the radiation-dominated era, the transfer function is defined as:
\begin{equation}
        {I^2}(t,s)  = 288\frac{(s^2 + t^2 - 6)^2}{(t^2 - s^2)^6}\bigg[ \frac{\pi^2}{4}(s^2 + t^2 - 6)^2\Theta(t - \sqrt{3}) + \left( {t^2 - s^2 - \frac{1}{2}(s^2 + t^2 - 6)\ln \left| \frac{t^2 - 3}{3 - s^2} \right|} \right)^2 \bigg]~,
        \label{I}
\end{equation}
Here, $\Theta$ is the Heaviside step function. The energy density fraction of gravitational waves is defined as:
\begin{equation}
    \Omega_{\mathrm{GW}}(k,\tau)\equiv\frac{\rho_{\mathrm{GW}}(k,\tau)}{\rho_{\mathrm{tot}}(\tau)}=\frac{1}{24}\left(\frac{k}{a(\tau)H(\tau)}\right)^2P_h(k,\tau)~.
\end{equation}
It is important to note that this quantity represents the relative energy density of gravitational waves compared to the total energy density of the universe, evaluated at conformal time $\tau$. However, for SIGW generated during radiation dominance, $\Omega_{\mathrm{GW}}$ can be related to the present-day observations by accounting for cosmic evolution. For $\rho_{\mathrm{GW}}=\Omega_{\mathrm{GW}}(k,\tau)\rho_{\mathrm{tot}}(\tau)\sim a^{-4}$, the present-day energy density fraction of SIGW is $\Omega^{\langle 0\rangle}_{\mathrm{GW}}\sim \Omega^{\langle 0\rangle}_rP^{\langle k\rangle}_h(k)/24$, where $\Omega^{\langle 0\rangle}_r\sim 10^{-5}h^{-2}$ is the present-day energy density fraction of radiation.
\begin{figure}
    \centering
    \includegraphics[width=0.75\linewidth]{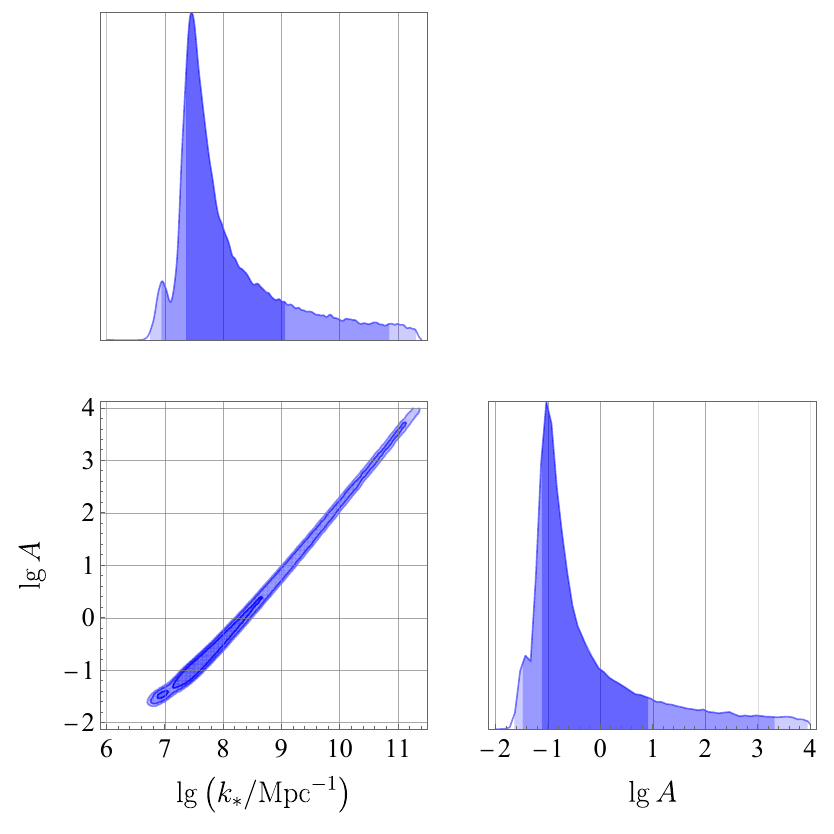}
    \caption{Posterior distributions of the BPL parameters $\{A, k_\ast\}$ from the constraints on SIGW by NANOGrav 15-yr data. Similar to \cite{Yang:2025vsq}, here from the dark to the light blue correspond to the $1\sigma$, $2\sigma$ and $3\sigma$ regions, respectively. The prior regions are set as $\lg A\in [-3, +4]$ and $\lg(k_\ast/\text{Mpc}^{-1})\in [5.8,12.8]$. }
    \label{fig:PosteriorBPL}
\end{figure}

\begin{figure}
    \centering
    \includegraphics[width=0.75\linewidth]{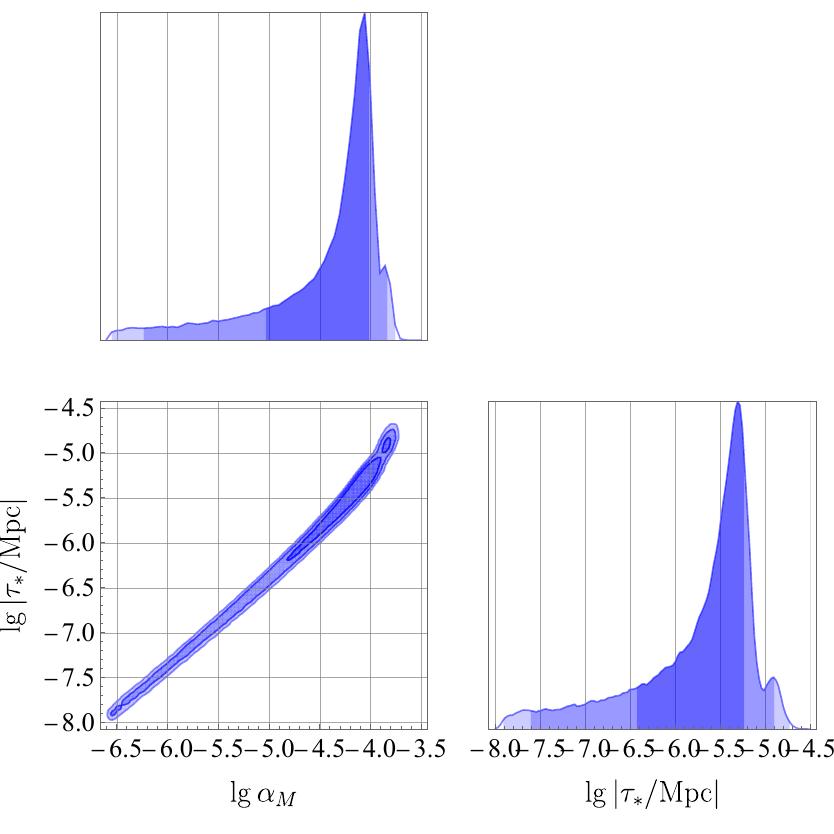}
    \caption{Posterior distributions of the model parameters $\{\alpha_M, \tau_\ast\}$ from the same constraints as above. Similarly, from the dark to light blue correspond to the $1\sigma$, $2\sigma$ and $3\sigma$ regions, respectively. }
    \label{fig:PosteriorMDR}
\end{figure}

In Figs.~\ref{fig:PosteriorBPL} and \ref{fig:PosteriorMDR} we plot the posterior distributions of BPL parameters $\{A, k_\ast\}$ as well as the model parameters $\{\alpha_M, \tau_\ast\}$ according to the constraints on the SIGW from the NANOGrav 15-yr data, using the public code \texttt{PTArcade}~\cite{Mitridate:2023oar}. We implemented the Markov chain Monte Carlo (MCMC) analysis, following the Gelman--Rubin convergence criteria $|R-1|\lesssim 0.01$ \cite{1177011136}. From Fig.~\ref{fig:PosteriorBPL} one can see that, the two parameters $A$ and $k_\ast$ are strongly positively correlated under such constraint, namely the larger the wavenumber of the power spectrum is, the larger the amplitude it will be. Very roughly speaking, this might be because that at the end of inflation, the fluctuation modes with larger number exit the horizon later and reenter earlier, therefore narrower window would be left for them to generate SIGW as well as PBHs. In order to generate the same amount of SIGW, their amplitudes should be enhanced accordingly. Moreover, the GW signal from the PTA data detected will set a lower bound to both $A$ and $k_\ast$, as shown in Fig.~\ref{fig:PosteriorBPL}.

Similarly, the MDR model parameters $\alpha_M$ and $\tau_\ast$ can also be constrained. From Eq.~\eqref{eq:MDR-BPL-log} or~\eqref{eq:MDR-BPL-inv}, these parameters are roughly anti-correlated with BPL parameters $A$ and $k_\ast$, respectively, therefore, there is also a positive correlation relationship between them, as shown in Fig.~\ref{fig:PosteriorMDR}. Moreover, upper bounds are set for both $\alpha_M$ and $\tau_\ast$. All the parameters behave the same as in Ref.~\cite{Yang:2025vsq}.

\subsection{PBH Abundance in the Compaction-Function Approach and Overproduction Avoidance}\label{TextOnCompactionFunction}

In \cite{Yang:2025vsq}, the abundance of primordial black holes (PBHs) is calculated using the compaction-function approach. In this approach, the key quantity is the compaction function ${\cal C}(\boldsymbol{x},r)$, which is related to both curvature perturbation $\zeta$ and its spatial derivative $\zeta^\prime$. To be precise, it is defined as \cite{Shibata:1999zs}:
\begin{equation}
    {\cal{C}}(\boldsymbol{x},r)\equiv 2\frac{\delta M(\boldsymbol{x},r)}{R(r)}~,
\end{equation}
where $\delta M$ is the mass excess within a sphere of areal radius $R$. This function measures the local density contrast and determines whether a region collapses to form a PBH. The mass fraction $\beta(r_H)$, representing the probability of PBH formation at the horizon scale $r_H$, is given by:
\begin{align}
    \label{MassFrac}
\beta(r_H) \equiv\frac{\rho_{\text{PBH}}}{\rho_{\text{tot}}}\Bigg|_{r_H} = \int_{\cal D} \frac{M_{\mathrm{PBH}}}{M_H} P(\zeta_G, {\cal C}_G)\mathrm{d}\zeta_G\,\mathrm{d}{\cal C}_G~,
\end{align}
with a mass of a PBH:
\begin{equation}
    \label{PBHmass}
    M_{\mathrm{PBH}}=\mathcal{K}({\cal C} - {\cal C}_{\mathrm{th}})^{\gamma_M}M_H~.
\end{equation}
Here, $\mathcal{K}$ and $\gamma_M$ are parameters depending on the equation of state, which have been set to be $\mathcal{K}=4$ and $\gamma_M=0.36$. $\mathcal{C}_{\mathrm{th}}$ is the threshold for collapse, and $P(\zeta_G, \mathcal{C}_G)$ is the joint probability distribution of the Gaussian curvature perturbation $\zeta_G$ and its derivative-related variable $\mathcal{C}_G$. The integral domain $\mathcal{D}$ requires $\mathcal{C} > \mathcal{C}_{\mathrm{th}}$ and the linear part $\mathcal{C}_l < 4/3$ to avoid type II perturbations. The total PBH abundance relative to dark matter is obtained by integrating over all horizon masses:
\begin{equation}
        \label{Abund}
        {f_{{\mathrm{PBH}}}}\equiv\frac{\rho_{\text{PBH}}}{\rho_{\text{DM}}}\Bigg|_{\text{form}}\simeq\frac{1}{\Omega_{\mathrm{DM}}}\int_{{M_{{H}}} = 0}^{ + \infty } {{{\mathrm{d}\ln}}{M_{{H}}}{{\left( {\frac{{{M_{{H}}}}}{{{2.8\times10^{17}M_\odot }}}} \right)}^{ - 1/2}}\beta}~,
\end{equation}
where $M_H$ is the horizon mass, and $\Omega_{\mathrm{DM}}$ is the dark matter density parameter. This formalism allows us to assess whether PBHs are overproduced when the SIGW spectrum is fitted to PTA data. If $f_{\mathrm{PBH}}$ exceeds unity, there will be an overproduction problem.

In \cite{Yang:2025vsq}, we numerically calculated the abundance of PBHs, using the compaction-function approach, showing that MDR can be helpful in resolving the overproduction problem. Without non-Gaussianities taken into account, the $f_{\mathrm{PBH}}=1$ line can be marginally included in the $2\sigma$ region allowed by the PTA data. Moreover, the negative non-Gaussianity will also play an important role. For $f_{\mathrm{nl}}=-1$, the $f_{\mathrm{PBH}}=1$ line will reach the edge of the $1\sigma$ region. We also calculate the mass distribution of the PBHs, in which PBHs can reach sub-solar masses of ${\cal O}(0.1\sim1)M_\odot$ under the model fitting the NANOGrav 15-yr data. In such a mass region, it is important to consider constraints from microlensing effect as well \cite{EROS-2:2006ryy, Niikura:2017zjd, Oguri:2017ock, Zumalacarregui:2017qqd, Niikura:2019kqi, Mroz:2024wia, Mroz:2024mse}. We also showed that with a little bit careful parameter choices, the abundance of the PBHs can also be made consistent with the microlensing constraints.

\section{Reanalysis Using Peak Theory}
\label{peak}

\subsection{The Peak Theory}
\label{PeakTheory}
Besides the compaction-function approach, another important approach of computing PBHs is the so-called peak theory \cite{Bardeen:1985tr}.  In peak theory,
it is hypothesized that PBHs form at density peaks exceeding a critical threshold, with the additional requirement that the PBH formation is exclusively associated with such peaks. Constraints can thus be imposed on PBHs at these peak locations; the corresponding peak number density (number of peaks per unit volume) is
\begin{equation}
    n(\mathbf{x})=\delta_D^{(3)}(\eta )\Theta _H(\lambda _3)\delta_D\left( \nu -\bar{\nu} \right) \Theta _H(\nu -\nu _{\mathrm{th}})\label{eq:number-density-peaks}~.
\end{equation}
In this expression, $\eta (\mathbf{x}) \equiv \sigma_{1}^{-1}\nabla \mathcal{C}(\mathbf{x})$ is the normalized gradient of the smoothed density contrast, $\delta_D^{(3)}$ selects the extreme points of the perturbation where $\eta=0$; $\Theta_{H}$ is the Heaviside step function, and $\lambda_{3}$ is the smallest eigenvalue of the second-derivative matrix of the smoothed density contrast, used to select the local maxima (i.e., peaks) and discard minima. The factor $\Theta_{H}(\nu - \nu_{\mathrm{th}})$ imposes a threshold on the perturbation amplitude.

It is interesting to see whether this approach can alleviate the overproduction problem as well. To compute the PBH abundance, one must average Eq.~\eqref{eq:number-density-peaks} over the parameter space, yielding the averaged number density of peaks as \cite{Bardeen:1985tr} (see also \cite{Yoo:2022mzl,Young:2024jsu}):
\begin{equation}
 \mathcal{N} \equiv \langle n\rangle= \int \mathrm{d}\mathbf{y}\,\frac{\sigma_{2}^{3}}{\sigma_{1}^{3}}\left|\operatorname{det}\zeta_{ij}\right|nP(\mathbf{y})~,
\end{equation}
where the parameter space is $\mathbf{y} = \left\{\nu = \frac{\mathcal{C}}{\sigma_{0}},\; \overrightarrow {\eta}= \frac{1}{\sigma_{1}}\nabla \mathcal{C}(\mathbf{x}),\; \zeta_{ij} = \frac{1}{\sigma_{2}}\frac{\partial^{2}\mathcal{C}(\mathbf{x})}{\partial x_{i}\partial x_{j}}\right\}$, and $P(\mathbf{y})$ is the multidimensional joint probability distribution function of these parameters.  Critical variances $\sigma_0$, $\sigma_1$ and $\sigma_2$ appearing in the PDF are defined as
\begin{align}
\sigma_n^2 &= \frac{16}{81}\int_0^\infty \frac{\mathrm{d}k}{k} (kr)^4 W^2(k,r)T^2(k,r)k^{2n} \mathcal{P}_\zeta \, \label{eq:sigma}
\end{align}
where
\begin{eqnarray}
    W(k,r) &=& 3\frac{\sin(kr) -kr\cos(kr)}{(kr)^3}~,\\
    T(k,r) &=& 3\frac{\sin(kr/\sqrt{3}) - (kr/\sqrt{3})\cos(kr/\sqrt{3})}{(kr/\sqrt{3})^3}
\end{eqnarray}
are the window function and the transfer function, respectively. Thus, the power spectrum, which is given in \eqref{BPL}, influences the PBH abundance through its effects on these variances.

Note that the parameter space has ten components, so the covariance matrix is ten-dimensional.  For better calculation, it is useful to perform a change of variables \cite{Bardeen:1985tr, Yoo:2022mzl, Young:2024jsu}:
\begin{equation}
  \begin{aligned}
u &= -\left(\zeta_{11}+\zeta_{22}+\zeta_{33}\right)/\sigma_2~, \\
v &= -\left(\zeta_{11}-\zeta_{33}\right)/2\sigma_2~, \\
w &= -\left(\zeta_{11}-2\zeta_{22}+\zeta_{33}\right)/2\sigma_2~.
\end{aligned}
\end{equation}
Under this transformation, we integrate the number density over the space of $\nu, u, v, w$. 
After integration, the peak number density becomes
\begin{equation}
\begin{aligned}
\label{Nnu}
   \mathcal{N}(\nu) = &\frac{\sigma_{2}^{3}}{3^{3/2}(2\pi)^{2}\sigma_{1}^{3}}\exp\left(-\frac{\nu^{2}}{2}\right)\Theta_{H}(\nu-\nu_{\mathrm{th}}) \\
&\times\int_{0}^{\infty}\mathrm{d}u\,f(u)\,\frac{1}{\sqrt{2\pi}(1-\gamma_{02}^{2})}\exp\left(-\frac{(u-\gamma_{02}\nu)^{2}}{2(1-\gamma_{02}^{2})}\right)~,
\end{aligned}
\end{equation}
where $\nu \equiv \mathcal{C}/\sigma_{0}$, $\gamma_{02} = \sigma_{1}^{2}/(\sigma_{0}\sigma_{2})$, and the function $f(u)$ is given by
\begin{equation}
\begin{aligned}
    f(u) = &\sqrt{\frac{2}{5\pi}}\left[\left(\frac{u^{2}}{2}-\frac{8}{5}\right)\exp\left(-\frac{5u}{2}\right)+\left(\frac{31u^{2}}{4}+\frac{8}{5}\right)\exp\left(-\frac{5u}{8}\right)\right] \\
&+\frac{1}{2}\left(u^{3}-3u\right)\left[\operatorname{erf}\left(\sqrt{\frac{5}{2}}u\right)+\operatorname{erf}\left(\sqrt{\frac{5}{8}}u\right)\right]~.
\end{aligned}
\end{equation}

Directly applying this method to compute the PBH abundance would be extremely cumbersome. Nevertheless, it is useful to apply the high-peak limit, where $u\gg 1$. Given that the formation of primordial black holes is very rare and occurs only from large-amplitude density perturbations, the peaks of interest are far above unity. Under this approximation, the function $f(u)$ can be simplified as \cite{Bardeen:1985tr, Yoo:2022mzl, Young:2024jsu}:
\begin{equation}
    f(u)\approx u^{3} - 3u\approx u^{3}~.
\end{equation}
Substituting this into Eq.~\eqref{Nnu} and changing the variable $\nu$ back to the smoothed density contrast $\mathcal{C}$, we obtain the simplified peak number density formula,
\begin{equation}
\label{NC}
    \mathcal{N}({\cal C}) = \frac{1}{3^{3/2}(2\pi)^{2}}\left(\frac{\sigma_{1}}{\sigma_{0}}\right)^{3}\left(\frac{\cal C}{\sigma_{0}}\right)^{3}P({\cal C}_G,\zeta_G)~.
\end{equation}

According to the probability function, we need to relate the smoothed density contrast $\mathcal{C}$ to its Gaussian component $\mathcal{C}_G$. For a radiation-dominated equation of state $w = 1/3$, one has
\begin{equation}
\mathcal{C} = \mathcal{C}_1(r) - \frac{3}{8}\mathcal{C}_1^2(r) \, . \label{eq:C_C1}
\end{equation}
where
\begin{equation}
    \label{comp4}
    {\cal C}_1(r)=\frac{4}{3}r\zeta_{,r}=\frac{4}{3}r\frac{\mathrm{d}\zeta}{\mathrm{d}\zeta_G}\zeta_{G{,r}}=\frac{4}{3}\frac{\mathrm{d}\zeta}{\mathrm{d}\zeta_G}{\cal C}_G~.
\end{equation}
Moreover, the probability distribution function $P(\zeta_G, {\cal C}_G)$ can be expressed as:
\begin{equation}
    \label{PDF}
P(\zeta_G, {\cal C}_G) = \frac{1}{2\pi\sqrt{|\bf\Sigma|}}\exp\left[{-\frac{1}{2}{{\bf{X}}^{\mathrm{T}}}{{\bf{\Sigma }}^{ - 1}}{\bf{X}}} \right]~,
\end{equation}
with variables matrix and covariance matrix
\begin{equation}
{\bf{X}}\equiv\left[ \begin{array}{l}
\zeta_G\\
{\cal C}_G
\end{array} \right],~~~{\bf{\Sigma}}\equiv\left[ {\begin{array}{*{20}{c}}
{\sigma _{\zeta}^2}&{\sigma _{\zeta{\cal C} }^2}\\
{\sigma _{\zeta{\cal C} }^2}&{\sigma_{\cal C} ^2}
\end{array}} \right]~,
\end{equation}
where the variances and the covariance $\sigma _{\zeta }$, $\sigma _{\zeta{\cal C} }$ and $\sigma _{{\cal C} }$ describing $\zeta$ and $\cal{C}$ are

\begin{align}
\sigma_\zeta^2 &= \int_0^\infty \frac{\mathrm{d}k}{k} (kr)^4 W^2(k,r)T^2(k,r) \mathcal{P}_\zeta \,,\\
\sigma _{\zeta{\cal C} }^2 &=\frac{4}{9} \int_0^\infty \frac{\mathrm{d}k}{k} (kr)^4 W(k,r)\;\mathrm{sinc}(kr)T^2(k,r) \mathcal{P}_\zeta \,,\\
\sigma _{\cal C}^2 &= \frac{16}{81}\int_0^\infty \frac{\mathrm{d}k}{k} (kr)^4 \;\mathrm{sinc}^2(kr)T^2(k,r) \mathcal{P}_\zeta \,
\label{eq:sigma2}
\end{align}

Note that Eq.~\eqref{NC} is the basic formula to calculate the PBH abundance using the peak theory.

\subsection{PBH Abundance}\label{PBH abundance}


In the peak theory approach, the PBH abundance can be obtained by integrating the peak number density above the threshold multiplied by the horizon volume,
\begin{equation}
\label{eq:beta_peak}
    \beta = V(R_{\mathrm{H}})\int_{\mathcal{C}_{\mathrm{th}}}^{\mathcal{C}_{\mathrm{max}}}\mathcal{K}(\mathcal{C} - \mathcal{C}_{\mathrm{th}})^{\gamma_M}\mathcal{N}(\mathcal{C})\,\mathrm{d}\mathcal{C}
\end{equation}
where the compaction function $\mathcal{C}$ has a maximum of $2/3$, corresponding to the upper limit of $\mathcal{C}_1=4/3$ when only the type-I perturbation is taken into account for the PBH generation. Moreover, the volume factor $V(r)$ depends on the smoothing window function. Under the top-hat window, we have $V(r) = 4\pi r^{3}/3$ \cite{Yoo:2022mzl,Young:2024jsu}.

Finally, integrating over all scales yields the total PBH abundance today, i.e., the fraction of dark matter,
\begin{equation}
    \begin{aligned}
    \label{fPBH}
f_{\mathrm{PBH}}
&= \frac{1}{\Omega_{\mathrm{DM}}}\int \frac{\mathrm{d}M_{\mathrm{H}}}{M_{\mathrm{H}}}\left(\frac{M_{\mathrm{eq}}}{M_{\mathrm{H}}}\right)^{1/2}\left(\frac{g_{*}}{106.75}\right)^{\frac{3}{4}}\left(\frac{g_{*s}}{106.75}\right)^{-1}\beta(M_{\mathrm{H}}).
\end{aligned}
\end{equation}



\begin{figure}
    \centering
    \includegraphics[width=1.0\linewidth]{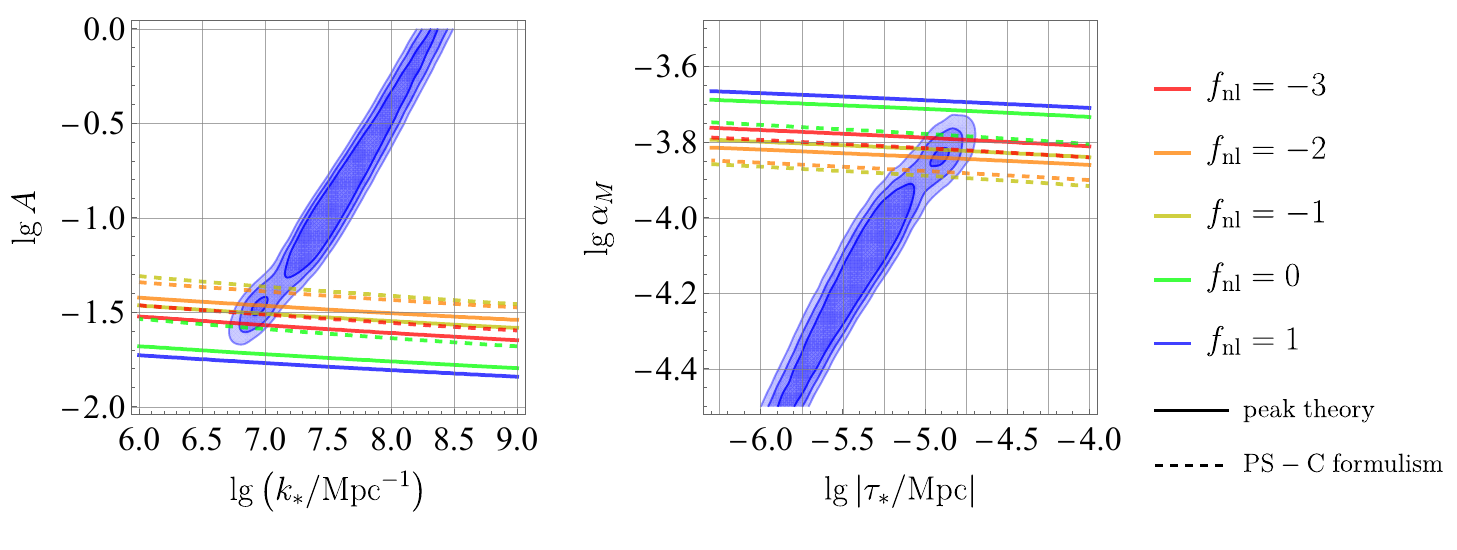}
    \caption{Posterior of the BPL (left) and MDR model (right) parameters from constraints on the NANOGrav 15-yr data, compared with $f_{\mathrm{PBH}}=1$ lines of the PBH abundance. The contours of the 2D posterior plot from dark to light correspond to the $1\sigma$, $2\sigma$, and $3\sigma$ confidence levels, respectively. The dashed and solid lines represent results obtained using the compaction function and the peak theory. Different colors mean that different values of $f_{\mathrm{nl}}$ have been taken into account, as shown beside the figure. For each case, the region above the line represents $f_{\mathrm{PBH}}>1$. }
    \label{PosteriorWithContour}
\end{figure}

In Fig.~\ref{PosteriorWithContour} we compare the posterior distribution for the BPL parameter $\{A, k_*\}$ as well as the MDR model parameter $\{\alpha_M, \tau_\ast\}$ obtained in Figs.~\ref{fig:PosteriorBPL} and~\ref{fig:PosteriorMDR}, to the $f_{\mathrm{PBH}}=1$ line of the PBH abundance, with different values of non-Gaussianity parameter $f_{\mathrm{nl}}$. We also compare the case where the PBHs are generated via the compaction-function approach (dashed lines) and the peak theory approach (solid lines). We see that, while non-Gaussianities have little effect on SIGW, they do have impacts on PBH generation, and will change the location of $f_{\mathrm{PBH}}=1$ line significantly. Fig.~\ref{PosteriorWithContour} shows that, for no non-Gaussianities involved in, i.e., $f_{\mathrm{nl}}=0$, while the $f_{\mathrm{PBH}}=1$ line for the compaction-function approach (the green dashed line) resides at the edge of the $2\sigma$ boundary, that for the peak theory approach (the green solid line) falls completely outside the $3\sigma$ region, meaning that the peak theory method will worsen the overproduction problem. This indicates that the peak theory will predict more PBHs compared to the compaction-function method. This is consistent with the result in Ref.~\cite{Young:2024jsu},  where the $\nu^3$ term is taken into account. Moreover, for a negative non-Gaussianity $f_{\mathrm{nl}}\leq -1$, the $f_{\mathrm{PBH}}=1$ lines for both approaches will reside well within the $2\sigma$ region, meaning that there will be no overproduction problem.


\subsection{PBH Abundance Distribution on Mass}
Besides the total fraction of PBHs in dark matter, it is also interesting to get the mass distribution of the PBHs, namely $f_{\mathrm{PBH}}(M_{\mathrm{PBH}})$, with $\int f_{\mathrm{PBH}}(M_{\mathrm{PBH}})\mathrm{d}\ln M_{\mathrm{PBH}}=f_{\mathrm{PBH}}$. To do this, one needs to perform variable substitution in the integrand from $\left( {{M_{{H}}},{{{\cal C}}_{{G}}},{\zeta _{{G}}}} \right)$ to $\left( {{M_{{\mathrm{PBH}}}},{M_{{H}}},{\zeta _{{G}}}} \right)$. Combining Eqs.~\eqref{PBHmass}, \eqref{eq:C_C1} and \eqref{comp4}, the relation between ${\cal{C}}_{G}$ and $M_\mathrm{PBH}$ can be represented as:
\begin{align}
    \label{CGofMPBH}
    {{{\cal C}}_{{G}}} = \frac{4}{3} {\left( {\frac{\mathrm{d}\zeta}{\mathrm{d}\zeta_G}} \right)^{ - 1}}\left[ {1 - \sqrt {1 - \frac{3}{2 }\left[ {{{{\cal C}}_{{\mathrm{th}}}} + {{\left( {\frac{{{M_{{\mathrm{PBH}}}}}}{{{\mathcal{K}}{M_{H}}}}} \right)}^{1/{\gamma_M} }}} \right]} } \right]~,
\end{align}
thus the mass distribution of PBHs is calculated as:
\begin{eqnarray}
    \label{fPBHofMPBH}
{f_{{\mathrm{PBH}}}}\left( {{M_\mathrm{PBH}}} \right)&=&\frac{1}{{{\Omega _{{\mathrm{DM}}}}}}\int_{{M_{{H}}^{\mathrm{min}}}\left( {{M_{{\mathrm{PBH}}}}} \right)}^{ + \infty }\bigg[\mathrm{d}\ln {M_{{H}}}{{\left( {2.8\times10^{17}M_\odot} \right)}^{1/2}}\nonumber\\
&\times& {M_\mathrm{PBH}^2} V(R_{\mathrm{H}})\int_\mathbb{R}^{} {\frac{\mathrm{d}{\cal C}_G}{\mathrm{d}M_\mathrm{PBH}}{\left(\frac{\mathrm{d}\zeta}{\mathrm{d}\zeta_G}\right)}^{-1}\mathrm{d}{\zeta _{{G}}}\mathcal{N}\left( {{{{\cal C}}_{{G}}},{\zeta _{{G}}}} \right)}\bigg]~,
\end{eqnarray}

\begin{figure}
    \centering
    \includegraphics[width=0.75\linewidth]{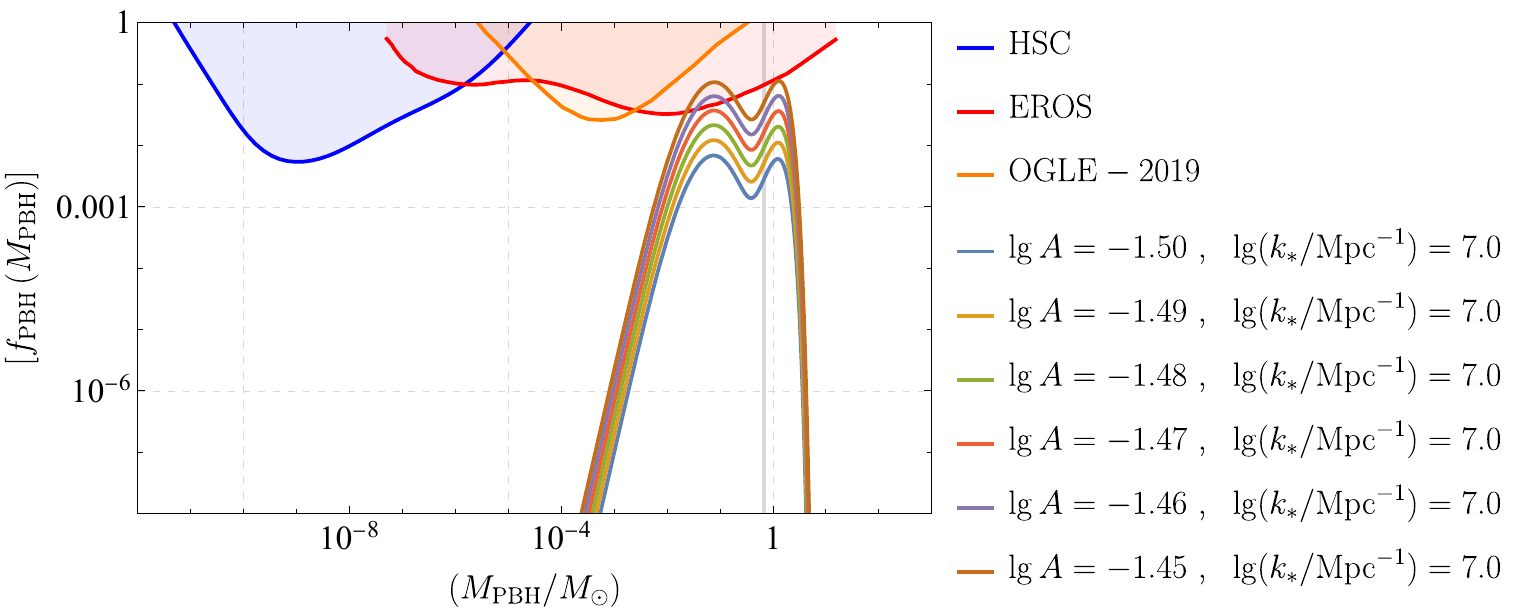}
    \caption{Mass fractions $f_{\mathrm{PBH}}(M_{\mathrm{PBH}})$ in logarithmic form, where $M_{\mathrm{PBH}}$ is normalized by $M_\odot$.  A series of curves with varying parameters of $\lg A$ are presented. The shaded regions show the constraints from microlensing observations by the EROS-2, the OGLE-2019 and the HSC surveys.}
    \label{fig:fPBHofMPBH}
\end{figure}

In Fig.~\ref{fig:fPBHofMPBH} we draw the logarithmic plot of $f_{\mathrm{PBH}}$ with respect to $M_{\mathrm{PBH}}$ (normalized by $M_\odot$), and take into account constraints from microlensing observations such as the EROS-2 \cite{EROS-2:2006ryy}, the OGLE-2019 \cite{Niikura:2019kqi} and the HSC \cite{Niikura:2017zjd} surveys. We plot a series of curves of $f_{\mathrm{PBH}}(M_{\mathrm{PBH}})$ with $\lg(k_\ast/{\mathrm{Mpc}}^{-1})$ fixed, while $\lg A$ varies between the value of $\lg A=-1.45$ and $\lg A=-1.50$. One can see that the same as in our previous paper \cite{Yang:2025vsq}, the PBHs generated here are mainly of sub-solar to solar masses. Moreover, for a little bit lower $\lg A$ (but still within the allowed region for overproduction avoidance), the curves can fit these observational constraints very well. 

\section{Conclusive Remarks}
\label{conclusion}
The problem of PBH overproduction has made the productivity of the PBH a sensitive issue. In this series of works, we considered the problem in the inflation model with a modified dispersion relation (MDR), which leads to a broken-power-law (BPL) curvature power spectrum. Our previous work addressed this issue using the compaction-function approach within the Press--Schechter (PS) formalism. However, there are various approaches for calculating the productivity of PBH, and different approaches may give different results. In this paper, we reanalyzed the same model using the peak theory approach, and in order to make the calculation feasible, we took the high-peak limit, which assumes that PBHs form only at rare, high-amplitude peaks of the curvature perturbation.

By analytical analysis and numerical calculation, we found that in the absence of the non-Gaussianity parameter $f_{\mathrm{nl}}$, although for the compaction-function approach we can alleviate the overproduction problem to the $2\sigma$ level, for the peak theory approach things will get worse. The $f_{\mathrm{PBH}}=1$ line falls outside the $3\sigma$ boundary of the allowed region from the PTA constraints. This is because in the perspective of peak theory, it tends to produce more PBHs than in the case of the compaction-function approach. In this sense, one would say that at least for this model, the peak theory approach of PBH generation might be less supported by the observational constraint. However, when small negative non-Gaussianity is involved in, say with $f_{\mathrm{nl}}\leq -1$, the $f_{\mathrm{PBH}}=1$ line will set off well within the $2\sigma$ region of the PTA constraint.

Moreover, we also consider the distribution of $f_{\mathrm{PBH}}$ in terms of the PBH mass, $M_{\mathrm{PBH}}$. Our numerical plot shows that in our case, PBHs are mainly distributed in the sub-solar-to-solar mass range, where the constraints from microlensing effects become important. However, we show that in the presence of small negative non-Gaussianities, it is easy to make the $f_{\mathrm{PBH}}(M_{\mathrm{PBH}})$ consistent with such constraints, by varying a little bit the parameters such as $\lg A$.

Some side remarks are in order. Recently PBHs and their generation have become hot topics; however, approaches of generation are not unique and in ongoing debate among one another. These approaches are based on different philosophies and scenarios, therefore, methodologically it is interesting and important to explore which approach is closer to the reality. Here in this paper, we discuss the two approaches of compaction function and the peak theory, and find their difference in the PBH abundance, by making use of the MDR inflation model for PBH generation. Although it is not an exhausted exploration, our results can make some contribution to distinguishing these two approaches. In the future it is alluring to make other tests to those approaches or find more optimistic ones, which will be implemented in an upcoming work.

\begin{acknowledgments}
We thank Xinzhe Zhang, Zhiyuan Yu, Zhengsheng Yang, Yuxuan Li and Zhi Yang for useful discussions. This work is supported by the National Natural Science Foundation of China (Grant No. 12575053) and the National Key Research and Development Program of China (Grant No. 2021YFC2203100).
\end{acknowledgments}

\bibliographystyle{apsrev4-2}
\bibliography{reference}

\end{document}